\documentclass[doublecol]{epl2} 

\usepackage{graphicx}
\usepackage{dcolumn}
\usepackage{bm}
\usepackage{float}
\usepackage{amsmath}
\usepackage{amssymb}
\usepackage{color}
\graphicspath{ {Images/} }

\title{Analysing degeneracies in networks spectra}
\shorttitle{Title} 

\author{Lo\"{i}c Marrec\inst{1,2} \and Sarika Jalan\inst{1,3}}

\institute{                    
  \inst{1} Complex Systems Lab, Discipline of Physics, Indian Institute of Technology Indore, Khandwa Road,
Simrol, Indore 453552, India\\
  \inst{2} Universit\'{e} Paris-Sud, 91405 Orsay Cedex, France \\
  \inst{3} Centre for Biosciences and Biomedical Engineering, Indian Institute of Technology Indore, Khandwa
Road, Simrol, Indore 453552, India
}
\pacs{89.75.-k}{Complex systems}
\pacs{02.10.Yn}{Matrix theory}

\abstract{
Many real-world networks exhibit a high degeneracy at few eigenvalues. We show that a simple transformation of the network's adjacency matrix provides an understanding to the origins of occurrence of high multiplicities in the networks spectra. We find that the eigenvectors associated with the degenerate eigenvalues shed light on the structures contributing to the degeneracy. Since these degeneracies are rarely observed in model graphs, we present results for various cancer networks. This approach gives an opportunity to search for structures contributing to degeneracy which might have an important role in a network. }

\begin{document}

\maketitle


The paper written by Leonhard Euler on the $Seven$ $Bridges$ $of$ $K\ddot{o}nigsberg$ marks a beginning of graph theory \cite{History} by introducing a concept of graphs representing complex systems. The work was restricted to small system size. Revolution in computing power later provided an opportunity to analyse very large real-world systems in terms of networks. Further, analysis of graph spectra has contributed significantly in our understanding of structural and dynamical properties of graphs \cite{Spectral_graph_theory1, Spectral_graph_theory2}.  Among other things, it has been noted that a symmetric spectrum about the origin is related to a bipartite graph \cite{Bipartite}. Further, bulk portion of eigenvalues has been shown to be modeled using random matrix theory \cite{rmt_network_sj}, whereas extremal eigenvalues have been shown to be modeled using the generalized eigenvalue statistics \cite{f,extreme2}. Recent investigations have revealed that real-world networks exhibit properties which are very different from those of the corresponding model graphs \cite{Aguiar,SJ_phya,Camellia_plos2014}. One of these properties is occurrence of degeneracy at $0$, $-1$ and $-2$ eigenvalues \cite{Spectral_graph_theory1}. Few papers have related $0$ and $-1$ eigenvalues to stars and cliques respectively \cite{Spectra_of_networks_1, Spectra_of_networks_3, Spectra_of_networks_4}. However, graphs in absence of stars and cliques can still show a degeneracy at $0$ and $-1$ eigenvalues, respectively. As a result, these reasons are not exhaustive and it turns out that origins of degeneracy at these eigenvalues are more complex. For example, the $0$ degeneracy has been shown to be resulted from the complete and the partial duplications \cite{Yadav_Chaos} of nodes which are particularly interesting for biological systems as they shed light on fundamental process in evolution related with gene duplication \cite{Duplication}, hence an interest lies in investigating origins of other degenerate eigenvalues. We will see in the following that two reasons emerge to explain degeneracy of every eigenvalue. In particular cases, one of these reasons reveals existence of characteristic structures in networks. 

In this paper, we consider finite undirected graphs defined by $ G=\{V,E\} $ with $V$ the node set, and $E$ the edge set such as $ \mid V \mid = N $ and $ \mid E \mid = m $. A graph is completely determined by its adjacency matrix for which its element $A_{ij}$ is $1$ when there is an edge from vertex $i$ to vertex $j$, and $0$ otherwise. {In the following, the rows $i$ of every adjacency matrix will be denoted by $R_i$}.



The eigenvalues are obtained by computing the roots of the characteristic polynomial of the adjacency matrix, $\chi_{A}(\lambda) = \mbox{det}(A - \lambda I) = \prod_{i=1}^{N}(\lambda-\lambda_{i})$ and denoted by $ \lambda_{1} \leq \lambda_{2} \leq \mbox{...} \leq \lambda_{N} $. Since the adjacency matrix of an undirected graph is symmetric with $0$ and $1$ entries, the eigenvalues are real. The associated eigenvectors \textbf{v}$_1$, \textbf{v}$_2$ ,..., \textbf{v}$_N$ satisfy the eigen-equation $A\textbf{v}_{i}=\lambda_{i}\textbf{v}_{i}$ with $i=1,2,...,N$.


A complete graph, denoted by $K$, is an undirected graph for which every pair of nodes is connected by a unique edge. This type of graphs is especially interesting since their spectra exhibit a very high multiplicity at $-1$ eigenvalue. Specifically, a complete graph of $N$ nodes has $N-1$ degeneracies for $-1$ eigenvalue \cite{Spectral_graph_theory1}. However, it is misleading to associate this special graph structure with $-1$ degeneracy. Let us take as example the 5 nodes complete graph in which we have removed an edge. In the resulting graph, two $-1$ eigenvalues are retained whereas the globally connected structure is destroyed, which indicates that the globally connected structure is not sufficient to explain occurrence of $-1$ degeneracy. We will see in the following that only one type of particular structure consisting of a complete graph and its variants contribute to $-1$ eigenvalue.  
 
We consider the matrix $ A + I $, where $ I $ is the identity matrix, and we make a change of variables in the characteristic polynomial such as $\chi_{A+I}(\lambda) = \chi_{A}(\mu)$. 
By this way, $ \mu $ is an eigenvalue of $A$ if and only if $ \lambda $ is an eigenvalue of $ A + I $. 
We can also prove that they have the same multiplicity. 
Hence, it is possible to reduce the computation of $-1$ eigenvalue of $ A $ to the $0$ eigenvalue of $ A + I $. This is especially interesting since the origin and implications of occurrence of $0$ degeneracy in networks spectra are well characterized \cite{Theorem_zero, Yadav_Chaos}. Spectrum of a matrix of size $ N $ and rank $ r $ contains $0$ eigenvalue with multiplicity $ N-r $. Three conditions lead to the lowering of the rank of a matrix; (i) if the network has
 an isolated node ($R_{i}=0 \cdots  0  \cdots \cdots 0$),  (ii)  At least two rows are equal ($R_{i}=R_{j}$), (iii) Two or more rows together are equal to some other rows ($\sum_{i} a_{i} R_{i}=\sum_{j} b_{j} R_{j}$,
where $a_{i}$ and $b_{i}$ take integer value included $0$).
In the case of $ A + I $, it is obvious that the condition (i) is never met.
We focus now on the condition (ii). Let us consider a network of size $ N $ for which two nodes labelled 1 and 2 verify $ R_{1}=R_{2} $ in the adjacency matrix added to the identity matrix. 

\begin{equation}
A+I  = 
 \begin{pmatrix}
  1 & a_{1,2} & \cdots & a_{1,N} \\
  a_{1,2} & 1 & \cdots & a_{2,N} \\
  \vdots  & \vdots  & \ddots & \vdots  \\
  a_{1,N} & a_{2,N} & \cdots & 1 
 \end{pmatrix}
\label{Eq16}
 \end{equation}

The condition (ii) is verified for any pair of rows, say $ 1^{st} $ and $ 2^{nd} $, if and only if $a_{1,2}=1$ and $a_{1,i}=a_{2,i}$ for $i=3,4,...,N$


So, the adjacency matrix $ A $ takes the following form :


\begin{equation}
A = 
  \begin{pmatrix}
     0 & 1 & \cdots & a_{1,N} \\
     1 & 0 & \cdots & a_{1,N} \\
     \vdots  & \vdots  & \ddots & \vdots  \\
     a_{1,N} & a_{1,N} & \cdots & 0  
  \end{pmatrix}
\label{Eq18}
\end{equation}

\begin{figure}[t]
\centerline{\includegraphics[width=0.8\columnwidth]{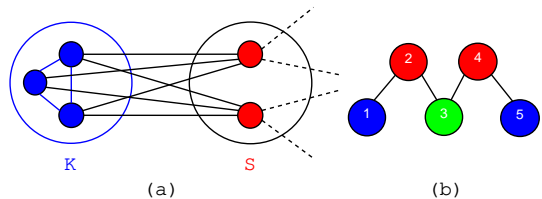}}
\caption{(Color online) \texttt{(a)} and \texttt{(b)} verify the condition (ii) and (iii) in $ A + I $, respectively.}
\label{Fig4}
\end{figure}

 The nodes $ 1 $ and $ 2 $ are interlinked and connected to the same set of other nodes. The rank of $ A + I $ is $ N-1 $, and hence we deduce that the spectrum associated to the $ A $ matrix contains exactly one $ -1 $ eigenvalue. It is trivial to generalize this proof to the case $R_{1}=R_{2}=...=R_{n}$. Hence, $n$ nodes forming a complete graph $ K $ connected to a same set $ S $ of other different nodes and denoted as $ K*S $ (Figure \ref{Fig4}) contribute to $-1$ eigenvalue with multiplicity $n-1$. 
 
 Next, we emphasize on the condition (iii), which for example may correspond to $ R_{1}+R_{2}=R_{3}+R_{4} $ in the matrix $ A + I $.  
%
%
For this particular case, the condition (iii) is satisfied if and only if $a_{1,j}+a_{2,j}=a_{3,j}+a_{4,j}$ for $j=1,2,...,N$.
%
%
On this basis, we can clarify that the condition (iii) implies for the node $j$ such as $j=5,6,...,N$. Whether the node $j$ is adjacent (respectively non-adjacent) to the both nodes labelled $1$ and $2$, it is also adjacent (respectively non-adjacent) to the both nodes $3$ and $4$. In the case where the node $j$ is either connected to $1$ or $2$, the condition (iii) imposes that $j$ is either connected to $3$ or $4$ (see Table \ref{table1}).

\begin{table}[h]
\centering
\begin{tabular}{c | c | c | c | c |}
 & Node 1 & Node 2 & Node 3 & Node 4 \\
\hline
 & 1 & 0 & 1 & 0 \\
\cline{2-5}
 & 1 & 0 & 0 & 1 \\
\cline{2-5}
Node $j$ & 0 & 1 & 1 & 0 \\
\cline{2-5}
 & 0 & 1 & 0 & 1 \\
\cline{2-5}
 & 1 & 1 & 1 & 1 \\
\cline{2-5}
 & 0 & 0 & 0 & 0 \\
 \hline 
\end{tabular}
\caption{Node $j$, such as $j=5,6,...,N$, is adjacent (respectively non-adjacent) to node $i$, such as $i=1,2,3$ and $4$, if the corresponding entry equals to 1 (respectively 0).}
\label{table1} 
\end{table} 

Let us now have a closer look at constraints which nodes 1, 2, 3 and 4 must obey. Since we have $a_{i,i}=1$, $a_{i,j}=a_{j,i}$ and by considering the previous constraints:
\begin{equation}
\left\{
    \begin{array}{ll}
        1+a_{1,2}=a_{1,3}+a_{1,4} \\
        a_{1,2}+1=a_{2,3}+a_{2,4} \\
        a_{1,3}+a_{2,3}=1+a_{3,4} \\
        a_{1,4}+a_{2,4}=a_{3,4}+1
    \end{array}
\right.
\label{Eq21}
\end{equation}

This set has more unknown variables than the number of equations. The system is underdetermined and has infinitely many solutions. As a result, it is difficult to define a typical structure which corresponds to the condition (iii). Here we will limit ourselves to illustrate it with the graph \texttt{(b)} in Figure \ref{Fig4} for which the adjacency matrix $A$ added to the identity matrix satisfies $R_{1}+R_{4}=R_{2}+R_{5}$.

This relation is at the origin of $-1$ eigenvalue observed in the spectrum of $A$. More generally, each linear combination of rows in $A+I$ leads to exactly one $-1$ eigenvalue.
The power of this approach is that it can be extended to all the degenerate eigenvalues. In the case of a network which exhibits a high multiplicity at the $x$ eigenvalue, it is wise to reduce the computation of $x$ eigenvalue to the study of 0 eigenvalue of $A-xI$ such as $\chi_{A-xI}(\lambda) = \chi_{A}(\mu)$.

In this manner, we are able to understand the origin of every degenerate eigenvalue, thus enabling to focus on their implications. We note that the condition (ii) is never met for $\lambda < -1$ and $0 < \lambda$ since the entries of the adjacency matrix are equal to $0$ or $1$.  

\begin{figure}
\centerline{\includegraphics[width=0.6\columnwidth]{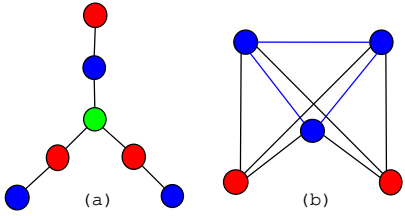}}
\caption{(Color online) \texttt{(a)} and \texttt{(b)} have same number of $-1$ eigenvalues but different structures. $-1$ degeneracy of \texttt{(a)} occurs through the condition (iii) whereas for \texttt{(b)} it results from the condition (ii).  }
\label{Fig6}
\end{figure}

  We have seen that $-1$ degeneracy in networks spectra is related to some typical structures. However, the study of eigenvalues and their multiplicities is not sufficient to determine the number and size of these structures in networks. For example, graphs of Figure \ref{Fig6} lead to the same number of $-1$ degeneracy but have different structures. The question we ask now is how can we identify nodes which contribute to degenerate eigenvalues? In order to address this, we consider eigenvectors associated to the degenerate eigenvalues of $A$. First we focus on $-1$ degeneracy and note that eigenvectors of $-1$ eigenvalue, such as $A\textbf{v}=\lambda_{-1}\textbf{v}$, are same as the eigenvectors corresponding to the $0$ eigenvalues of $A+I$ which verify $(A+I)\textbf{v}=\lambda_{0}\textbf{v}$.

We find that most of the entries of such eigenvectors (corresponding to $0$ eigenvalue) are equal to zero. 
It turns out that non-null entries reveal nodes which contribute to decreasing the rank of a matrix. Moreover, nodes belonging to the same structure, say $K*S$, are linked by the following 
relation (derivation is in the Supplementary Material \cite{SM}): 

\begin{equation}
\left\{
    \begin{array}{ll}
	\sum_{i\in K_p} v_{i}=0 \mbox{ with } v_{i}\neq0 \mbox{ and }p=1,2,...,n_{K*S}  \\
	v_{j \in V\setminus \{K_1\cup K_2 \cup ... \cup K_{n_{K*S}}\}}=0
    \end{array}
\right. 
\label{Eq32} 
\end{equation}

{where $n_{K*S}$ is the number of $K*S$ structures in the network}. This relation, arising directly from $R_{i}=R_{j}$, enables us to distinguish each structure contributing to $-1$ degeneracy through condition (ii). The same reason holds good for condition (iii). Indeed, nodes belonging to a same linear combination verify: 






 \begin{equation}
 \left\{
    \begin{array}{ll}
	\sum_{i\in (LC)_p} v_{i}=0 \mbox{ with } v_{i}\neq0  \mbox{ and }p=1,2,...,n_{LC}\\
    v_{j \in V\setminus \{(LC)_1\cup (LC)_2 \cup ... \cup (LC)_{n_{LC}}\}}=0 
    \end{array}
\right.
\label{Eq36} 
\end{equation}

{where $n_{LC}$ is the number of linear combinations of rows in the network}. The derivation of Eq. \ref{Eq36} follows the same reasoning as this of Eq. \ref{Eq32} seeing that $R_{k}=R_{l}$ is a particular case of $\sum_{i}a_{i}R_{i}=\sum_{j}b_{j}R_{j}$. It is also interesting to note that Eqs. \ref{Eq32} and \ref{Eq36} can be proved using row equivalent forms of adjacency matrices \cite{SM}. {Since the assumptions are only based on the conditions (ii) and (iii), the proofs provided in the Supplementary Material show that Eqs. \ref{Eq32} and \ref{Eq36} are valid for any arbitrary network.} These properties give the opportunity to find in every network the nodes which contribute to $-1$ degeneracy. We can go further by distinguishing the nodes which satisfy the condition (ii) and those which satisfy the condition (iii). In order to do it, we can have a rather easy computation of the rows of $A+I$ such as $R_{i}=R_{j}$. Then, by considering one of the eigenvectors associated to $-1$ eigenvalue, we can find all the non-zero entries. Among these non-zero entries, those which do not correspond to the nodes computed previously, satisfy the condition (iii). Besides being able to find the nodes leading to $-1$ degeneracy, we can associate them to (ii) or (iii).
As we did in the previous section, we extend this approach to all the degenerate eigenvalues. Indeed, eigenvectors of $x$ of $A$ matrix which satisfy $(A-xI)\textbf{v}=\lambda_{0}\textbf{v}$ are same as the eigenvectors corresponding to $0$ eigenvalues of $A-xI$ such as $A\textbf{v}=\lambda_{x}\textbf{v}$.





So, we can find precisely the origin of every degenerate eigenvalue, namely the condition (ii) or the condition (iii). In addition, we are able to identify the nodes which contribute to high multiplicity of eigenvalues. In brief, this approach provides a quantitative measure of degeneracy in networks spectra. 

\vspace{0.25cm}
What we have done so far is finding subgraphs behind occurrence of degeneracy. The question we ask now is: do they play a significant role in real-world networks? In order to answer this, we are going to assess whether randomness enables to observe this kind of structures using various model graphs. The $0$ degeneracy being the subject of a previous discussion \cite{Yadav_Chaos}, we focus in the following on degeneracy at $-1$ eigenvalue.  First, we consider Erd\"{o}s-Renyi model (ER) \cite{ER} in which each edge has a probability $p$ of existing. Since the edges are placed randomly, most of the nodes have a degree close to the average degree $\langle k \rangle$ of the graph \cite{ER}. So, the probability $p$ equals to $\dfrac{\langle k \rangle}{N}$.

\begin{figure}
\centerline{\includegraphics[width=\columnwidth]{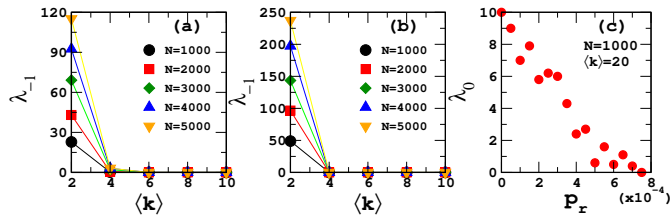}}
   \caption{(Color online) \texttt{(a)} and \texttt{(b)} illustrate the effect of the average degree $\langle k \rangle$ on $-1$ degeneracy for different sizes $N$ in ER and SF networks, respectively. All values are averaged over 20 random realizations of the network. \texttt{(c)} represents the number of 0 eigenvalues depending on the link rewiring probability for a 1000 nodes small-world network. The regular graph has degree $k=20$. All values are averaged over 20 random realizations of the network.}
   \label{Plot}
\end{figure}

We note that such networks are almost surely disconnected if $p<\frac{\ln(N)}{N}$ \cite{ER}.
In other words, if the previous equation is verified, there are at least two nodes such that no path has them as endpoints.
We construct ER networks with different average degrees $\langle k \rangle$ and sizes $N$. As depicted by Figure \ref{Plot}, ER networks exhibit a low degeneracy for small average degrees. We can explain this by referring to $n$-complete graphs percolation \cite{Clique_percolation}. Indeed, it has been shown that in such graphs, complete subgraphs occur beyond a certain probability $p$. The threshold of this percolation is defined as $p_{c}(n)=[(n-1)N]^{-\dfrac{1}{n-1}}$ \cite{Clique_percolation}. Higher the $n$, more is the increase in the threshold of percolation. Put another way, the more $n$ augments, the more the probability to have $n$-complete graphs in the ER graphs gets diminished. 
Because $p=\dfrac{\langle k \rangle}{N}$ and by referring to $p_{c}$, it is very easy to show that $n$-complete subgraphs appear in a network if the average degree is more than $\langle k \rangle_{c} \sim N^{\frac{n-2}{n-1}}$ \cite{Cliques_SF}. For example, $\langle k \rangle_{c}$ $\sim$ $31.62$ for $n=3$ and $N=1000$. In the case where $\langle k \rangle$ is less than $\langle k \rangle_{c}$, we do not observe $n$-complete subgraphs in ER networks. So, the structure $K*S$ being not obtained, the condition (ii) is not met. Whether $\langle k \rangle_{c}$ is less than $\langle k \rangle$, it is possible to find cliques in the networks. However, randomness which characterizes the ER model may not give the opportunity to organized structures such as $K*S$ to emerge. Nevertheless, the previous explanations are not exhaustive since the condition (iii) can contribute to $-1$ degeneracy. The condition not corresponding to a defined structure, it is more difficult to predict its contribution. Finally, the alert reader may inquire why we observe degeneracies for $\langle k \rangle=2$. As it is specified earlier, in the case where $p<\frac{\ln(N)}{N}$ is verified, the network is most probably disconnected. More particularly, for a given $N$, lower the value of $p$ (and hence $\langle k \rangle$), higher is the number of isolated nodes and two connected nodes \cite{ER}. Two connected nodes is a $K*S$ structure where $K$ contains two nodes and $S$ contains no node. That is why -1 degeneracy is existing for the low average degree.

We focus now on scale-free (SF) networks \cite{SF}. This model is particularly interesting since its degree distribution follows a power law as also observed for many real-world networks \cite{ER}. We generate the SF networks by using a preferential attachment process. Each new node is attached to the existing nodes with a probability which is proportional to their degrees. As a result of the power law, SF networks contain a few high degree nodes and a large number of low degree nodes and $-1$ degeneracy is not observed in most of the cases (Figure \ref{Plot}). As explained in \cite{Yadav_Chaos}, any two nodes having a low degree are more susceptible to connect to the high degree nodes, leading to two nodes having the same neighbors. However, the preferential attachment makes it less probable for these nodes to be interlinked. So, it is unlikely to find $K*S$ structure in SF networks. In the case of $\langle k \rangle=2$, we find $-1$ degeneracy. Thanks to the eigenvectors analysis, we observe that only the condition (iii) contributes to high multiplicity in this particular case. As we have already said, there is no typical structure corresponding to this condition. Therefore, it is difficult to provide an explanation to this observation. 
\begin{table}[t]
\centering
\begin{tabular}{c c c c c c c} 
\hline
$k$ & 10 & 20 & 30 & 40 & 50 & 60 \\
\hline
$\lambda_{0}$ & 5 & 10 & 5 & 20 & 25 & 5\\
\hline
$\lambda_{-1}$ & 0 & 0 & 0 & 0 & 0 & 0 \\
\hline
\end{tabular}
\caption{$0$ and $-1$ degeneracy in small-world network without link rewiring such as $N=1000$. {Multiplicity of $\lambda$ eigenvalue equals to the number of times that $n_m^{\lambda}$ is an integer, for $m=1,2,...N$}.}
\label{table6} 
\end{table} 
%
%
\begin{center}
\begin{table*}[t]
{\small
\hfill{}
\begin{tabular}{c c c c c c c c c c} 
\hline
Network & $N$ & $\langle k \rangle$ & $p$ & $p_{c}(3)$ & $\lambda_{+1}$ & $\lambda_{0}$ & $\lambda_{-1}$ & $<C.C>$ & $r$ \\
\hline
Breast$_{N}$ & 2443 & 12.38 & 0.0051 & 0.0143 & 0 & 72 & 21 & 0.28 & 0.08 \\
Breast$_{D}$ & 2046 & 13.83 & 0.0068 & 0.0156 & 0 & 71 & 12 & 0.29 & 0.19 \\
\hline
Colon$_{N}$ & 4849 & 16.05 & 0.0033 & 0.0102 & 0 & 164 & 19 & 0.25 & 0.18 \\
Colon$_{D}$ & 3423 & 21.23 & 0.0062 & 0.0121 & 0 & 44 & 10 & 0.23 & 0.09 \\
\hline
Oral$_{N}$ & 2105 & 20.66 & 0.0098 & 0.0154 & 0 & 60 & 13 & 0.31 & 0.19 \\
Oral$_{D}$ & 1542 & 34.75 & 0.0225 & 0.0180 & 0 & 15 & 1 & 0.35 & -0.03 \\
\hline
Ovarian$_{N}$ & 1748 & 7.77 & 0.0044 & 0.0169 & 4 & 129 & 31 & 0.25 & -0.01 \\
Ovarian$_{D}$ & 2022 & 7.95 & 0.0039 & 0.0157 & 2 & 116 & 19 & 0.26 & 0.10 \\
\hline
Prostate$_{N}$ & 2304 & 9.57 & 0.0042 & 0.0147 & 2 & 125 & 47 & 0.29 & 0.08 \\
Prostate$_{D}$ & 4938 & 7.62 & 0.0015 & 0.0101 & 4 & 340 & 135 & 0.30 & 0.10 \\
\hline
\end{tabular}}
\hfill{}
\caption{Statistical properties for all the normal and disease networks. The total number of nodes is $N$, the average degree $\langle k \rangle$, the probability $p$ of two nodes to be connected by an edge and the threshold of $3$-complete percolation $p_{c}(3)$. $\lambda$ represent the number of eigenvalues in the cancer networks, $<C.C>$ the average clustering coefficient and $r$ the assortativity value.}
\label{table2}
\end{table*}
\end{center}
%
As we have just seen, -1 degeneracy is not observed in ER and SF networks, leading us to believe that randomness is not conducive to degeneracy in networks spectra. In order to convince us, we study small-world (SW) networks constructed with the Watts-Strogatz mechanism \cite{SW}. This graph model is interesting since it illustrates small-world phenomenon according to which distance between nodes increases as the logarithm of the number of nodes in the network \cite{SW_properties}. The generating mechanism consists of a regular ring lattice where each node is connected to $k$ neighbors and for which edges are rewired with probability $p_{r}$. {The spectrum $\{\lambda_1, \lambda_2,...,\lambda_N\}$} of the Watts-Strogatz graph without a link rewiring, i.e. 1-d lattice with circular boundary 
condition, can be computed by $\displaystyle \lambda_{m}=\frac{\sin(\frac{\pi(N-(m-1))(k+1)}{N})}{\sin(\frac{\pi(N-(m-1))}{N})}-1$ with $m=1,2,...,N$. This relation leads to $\lambda=-1$ if and only if {the following constraint is fulfilled}:



\begin{equation}
\left\{
    \begin{array}{ll}
	n_{m}^{\lambda_{-1}}=\frac{(N-(m-1))(k+1)}{N} \\
	q_{m}^{\lambda_{-1}}\neq\frac{N-(m-1)}{N} \\
    \end{array}
\right. 
	\mbox{with } n_{m}^{\lambda_{-1}} \mbox{ and } q_{m}^{\lambda_{-1}} \mbox{ integers}
\label{Eq43}
\end{equation}

{In other terms, if $n_{m}^{\lambda_{-1}}$ is an integer and $q_{m}^{\lambda_{-1}}$ different as an integer, then $\lambda_m=-1$"}. The same reasoning applied in the case of $0$ eigenvalue which leads to:

\begin{equation}
n_{m}^{\lambda_{0}}=\frac{k}{2}\frac{N-(m-1)}{N} \mbox{ with } n_{m}^{\lambda_{0}} \mbox{ integer}
\label{Eq44}
\end{equation}

Thanks to these equations, it is possible to predict the multiplicity of $0$ and $-1$ eigenvalues. We compute $n_{m}^{\lambda}$ for every value of $m$, then the number of times where $n_{m}^{\lambda}$ is an integer equals to the multiplicity of $\lambda$. In the particular case of $n_{m}^{\lambda_{-1}}$, we ensure that we do not count it when $q_{m}^{\lambda_{-1}}$ is an integer.

As reported in Table \ref{table6}, there is no $-1$ degeneracy in the network for $p_{r}=0$. Mathematically, it may be due to the supplement constraint that $\lambda_{m}$ must verify to equal $-1$. However, we observe a degeneracy at $0$ eigenvalue. It is interesting to notice that the degeneracy does not increase constantly with an increase in $k$. Let us now how attempt to understand the probability $p_{r}$ affects the multiplicity of $0$ eigenvalue for a small-world network.  
Figure \ref{Plot} reveals that the number of $0$ eigenvalues decreases quickly with $p_{r}$. More particularly, $0$ degeneracy is completely removed for low link rewiring probability. Simulations for different configurations ($N$, $k$) yield similar results. As a consequence, the introduction of randomness, even small, have strong impacts on the multiplicity of eigenvalues.


\vspace{0.25cm}
We substantiate the previous results by considering examples of few real-world networks. We analyse protein-protein interaction networks (PPI) of five cancers namely Breast, Colon, Oral, Ovarian and Prostate \cite{Aparna_paper}. The PPI networks have proteins as nodes and the interaction between those proteins as edges.
These networks exhibit $0$ and $-1$ degeneracies (see Table \ref{table2}). In addition, we find a low degeneracy at $+1$ in few of these networks. The number of $0$ eigenvalues being more than the number of $-1$ eigenvalues indicates that duplication structures are more frequent than the $K*S$ structures. {
It may be tempting to make a causal link between macroscopic properties such as clustering coefficient or assortativity and eigenvalue degeneracy. However, as reported in
Table 3, the normal Oral cancer network has $\langle CC \rangle=0.31$ and $N_{\lambda_0}=60$
whereas for the disease one, $\langle CC \rangle=0.35$ and $N_{\lambda_0}=15$. What follows that
one can find networks having lesser degeneracy but with high value of $\langle CC \rangle$.
Furthermore, let us consider one more macroscopic quantity which is degree-degree correlations
and let us attempt to find a relation between assortatvity (positive degree-degree correlations and
degeneracy at -1). Again, let us consider the Ovarian cancer which for the disease case has
 $r=0.10$ and $N_{\lambda_0}=116$, whereas for the normal case has $r=-0.01$ and $N_{\lambda_0}=129$. Therefore, even if assortativities of networks are different, degeneracies may be quite the same.
All these point out that there exists no obvious causal link between these macroscopic properties and degeneracy in networks. Our experiments only indicates that randomness has a strong impact on degeneracy in networks spectra.}

Thanks to Eq. \ref{Eq32} and \ref{Eq36}, we can go further by using what we know about eigenvectors  associated  to $-1$  eigenvalue  in order to identify each contribution to degeneracy in  the  cancer networks. Table \ref{table3} reports the number of $-1$ eigenvalues, {denoted by $\lambda_{-1}$}, and nodes, {denoted by $N_{\lambda_{-1}}$}, by condition for all the normal and disease networks. The condition (ii) is largely at the origin of $-1$ degeneracy and in a few cases, condition (iii) is not met. Therefore, $K*S$ structure mostly contributes to $-1$ degeneracy in the networks spectra. As a conclusion, real-world networks contain a large number of $K*S$ structures whose existence is revealed by high multiplicity of $-1$ eigenvalues. As degeneracy at $-1$ eigenvalue is poorly observed in model networks, this indicates that the resulting structures may have a significance in real-world networks.

\begin{table}[t]
\centering
\begin{tabular}{c c c c c} 
\hline
Network & $\lambda_{-1}^{(ii)}$ & $\lambda_{-1}^{(iii)}$ & $N_{\lambda_{-1}}^{(ii)}$ & $N_{\lambda_{-1}}^{(iii)}$  \\
\hline
Breast$_{N}$ & 19 & 2 & 38 & 14 \\
Breast$_{D}$ & 12 & 0 & 21 & 0 \\
\hline
Colon$_{N}$ & 18 & 1 & 32 & 6 \\
Colon$_{D}$ & 10 & 0 & 19 & 0 \\
\hline
Oral$_{N}$ & 13 & 0 & 21 & 0 \\
Oral$_{D}$ & 1 & 0 & 2 & 0 \\
\hline
Ovarian$_{N}$ & 23 & 8 & 40 & 36 \\
Ovarian$_{D}$ & 13 & 6 & 24 & 28 \\
\hline
Prostate$_{N}$ & 34 & 13 & 56 & 58 \\
Prostate$_{D}$ & 117 & 18 & 199 & 91 \\
\hline
\end{tabular}
\caption{Number of $-1$ eigenvalues and nodes by condition, {namely the condition (ii) and (iii), for all the normal and disease} networks. $\lambda_{-1}$ and $N_{\lambda_{-1}}$ are the number of $-1$ eigenvalues and nodes respectively.}
\label{table3} 
\end{table}

\vspace{0.25cm}
As a conclusion, many real-world networks exhibit a very high degeneracy at few eigenvalues such as $0$ and $-1$ as compared to their corresponding random networks. This suggests that the nodes contributing to high multiplicity may play a central role in these networks. This {Letter} has numerically and analytically demonstrated the origin as well as structures contributing to degeneracy, giving the opportunity to study their impact in real-world networks. In the case of cancer networks, if such structures turn out to have a biological significance, the proposed approach will provide a new and different way to search for drug targets and biomarkers \cite{Conclusion}.

\acknowledgments
SJ is grateful to Department of Science and Technology, Government of India grant EMR/2014/000368 for financial support. LM acknowledges Sanjiv K. Dwivedi, Alok Yadav and Aparna Rai for help with 
the spectral analysis and cancer data, respectively.

\end{document}


\title{Supplementary material for \\ ''Analysing degeneracies in networks spectra''}

\author{{Lo\"{i}c Marrec$^{1,2}$ and Sarika Jalan $^{1,3}$ \email{sarika@iiti.ac.in} }\\
{\small \em $^1 $ Complex Systems Lab, Indian Institute of Technology Indore, Simrol Campus, Khandwa Road, Simrol, Indore 453552, India\\
$^2$ Universit\'{e} Paris-Sud, 91405 Orsay Cedex, France\\
$^3$ Centre for Biosciences and Biomedical Engineering, Indian Institute of Technology Indore, Simrol Campus, Khandwa Road, Simrol, Indore 453552, India}}
\date{\today}
\maketitle

This Supplementary Material is meant solely to provide the reader with an understanding of "Analysing degeneracies in complex networks" thanks to examples. Then, we propose a derivation of the relation which links nodes belonging to a same $K*S$ structure, namely:

\begin{equation}
\left\{
    \begin{array}{ll}
	\sum_{i\in K_p} v_{i}=0 \mbox{ with } v_{i}\neq0 \mbox{ and }p=1,2,...,n_{K*S}  \\
	v_{j \in V\setminus \{K_1\cup K_2 \cup ... \cup K_{n_{K*S}}\}}=0
    \end{array}
\right.
\label{Relation} 
\end{equation}

We also focus on the relation verified by nodes belonging to a linear combination, namely:

\begin{equation}
\left\{
    \begin{array}{ll}
	\sum_{i\in (L.C)_p} v_{i}=0 \mbox{ with } v_{i}\neq0  \mbox{ and }p=1,2,...,n_{L.C}\\
    v_{j \in V\setminus \{(L.C)_1\cup (L.C)_2 \cup ... \cup (L.C)_{n_{L.C}}\}}=0 
    \end{array}
\right.
\label{Relation2} 
\end{equation}

\section{Useful examples of graphs}

\begin{figure}[h]
\centerline{\includegraphics[width=0.4\textwidth]{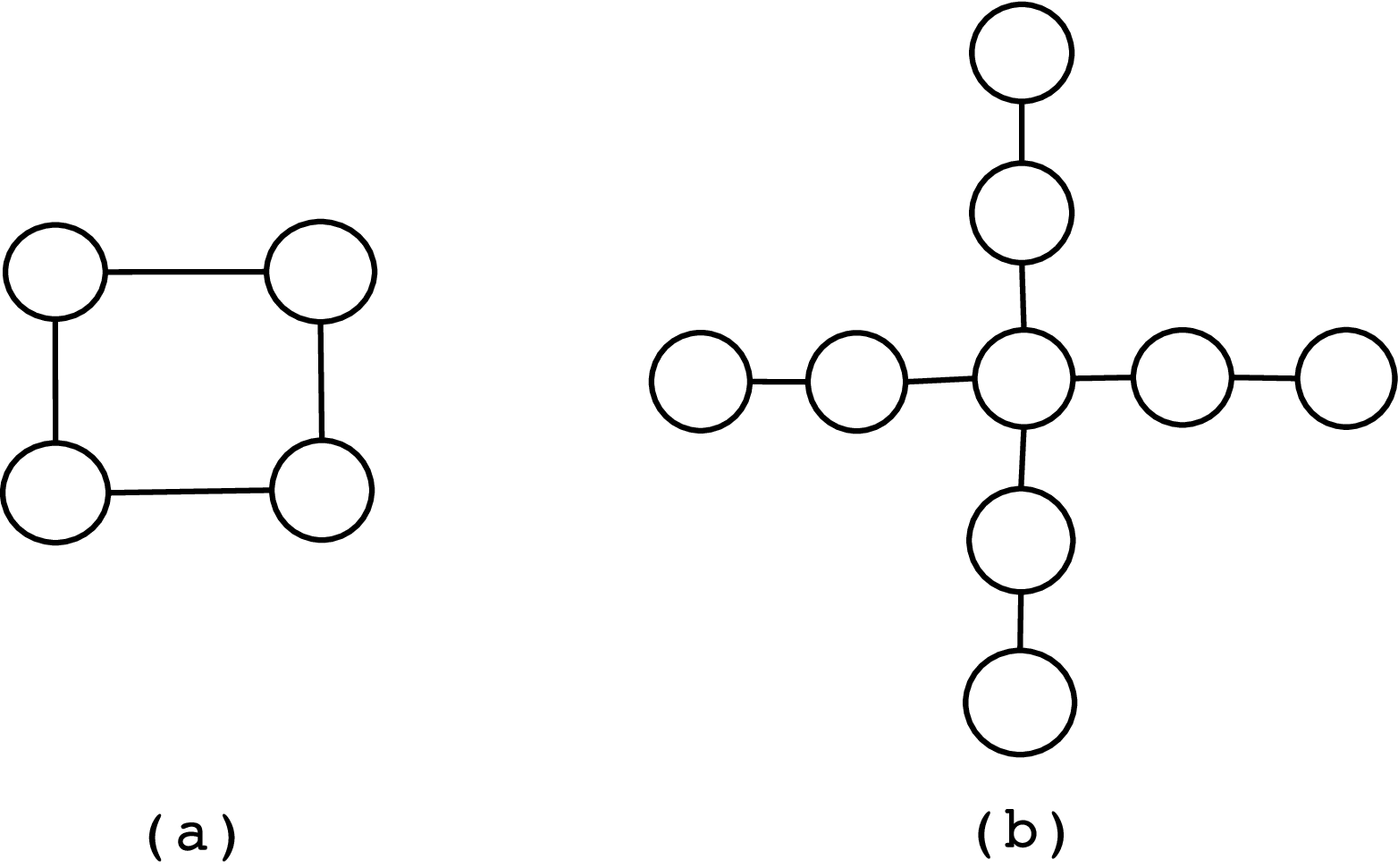}}
\caption{\texttt{(a)} has degeneracy at $0$ but no stars and \texttt{(b)} has degeneracy at $-1$ but no cliques.}
\label{Stars_cliques}
\end{figure}

\begin{figure}[h]
\centerline{\includegraphics[width=0.5\textwidth]{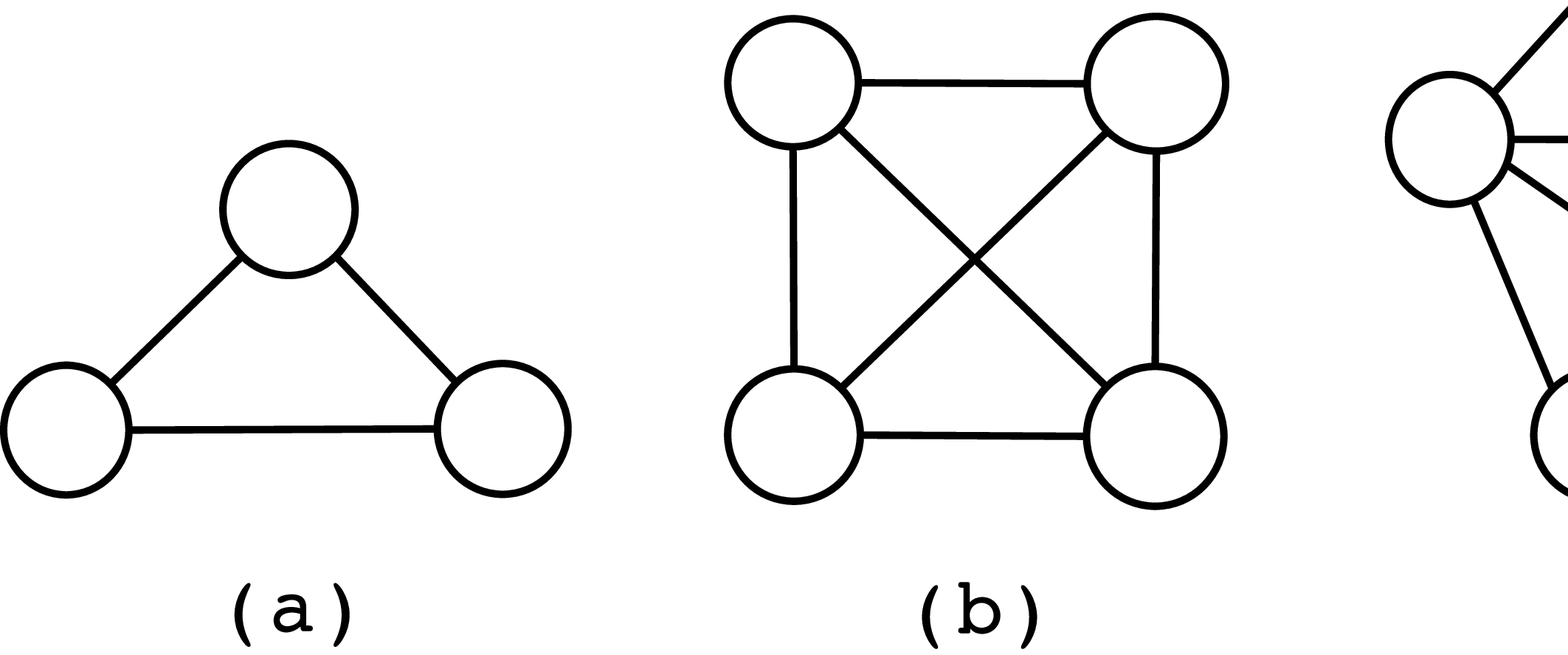}}
\caption{Complete graphs for different sizes. \texttt{(a)}, \texttt{(b)} and \texttt{(c)} have two, three and four $-1$ eigenvalues, respectively.}
\label{Cliques}
\end{figure}

\begin{figure}[h]
\centerline{\includegraphics[width=0.5\textwidth]{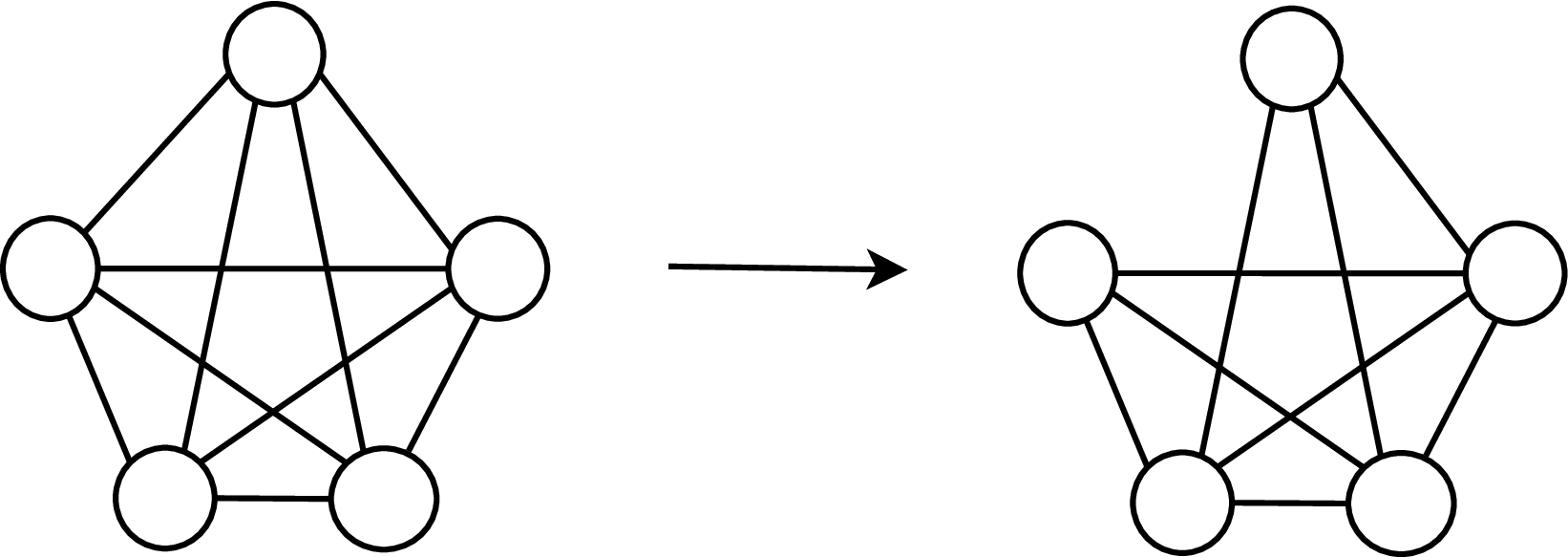}}
\caption{Complete graph of five nodes for which we cut one edge. The resulting graph exhibits two $-1$ eigenvalues.}
\label{Cut_edge}
\end{figure}

We mentioned in the manuscript that stars and cliques are not sufficient to explain the occurrence of $0$ and $-1$ eigenvalues, respectively. Graphs in Figure \ref{Stars_cliques} enable to illustrate this point. In order to go further, we focus on cliques. They consist of structures for which every pair of nodes is connected by a unique edge. Few examples are represented in Figure \ref{Cliques} for different number of nodes. Typically, the spectra of such graphs exhibit $N-1$ degeneracies for $-1$ eigenvalue, with $N$ the number of nodes. However, if we take the third graph of Figure \ref{Cliques} and we cut one of its edges, we will see that two $-1$ eigenvalues are retained whereas the globally connected structure is destroyed. (Figure \ref{Cut_edge}) This kind of observations motivated to search for other reasons behind occurrence of degeneracy in networks spectra.

\section{Derivation of Eq (\ref{Relation})}

In order to prove the equation (\ref{Relation}), we consider the eigen-equation:

\begin{equation}
(A+I)\textbf{v}=0
\end{equation}

In the following, we take $R_{1}=R_{2}$. It will be easy to extend to $R_{1}=R_{2}=...=R_{n}$.

\begin{equation}
  \begin{pmatrix}
     1 & 1 & \cdots & a_{1,N} \\
     1 & 1 & \cdots & a_{1,N} \\
     \vdots  & \vdots  & \ddots & \vdots  \\
     a_{1,N} & a_{1,N} & \cdots & 1  
  \end{pmatrix}
   \begin{pmatrix}
  v_{1} \\
  v_{2} \\
  \vdots \\
  v_{N}
 \end{pmatrix}
 =0
\label{EQ2}
\end{equation}

Thanks to the two first rows, we can write:

\begin{equation}
v_{1}+v_{2}+\sum_{i=3}^{N}a_{1i}v_{i}=0
\label{A}
\end{equation}

By extending to the general case, we obtain:

\begin{equation}
\sum_{i\in K}v_{i}+\sum_{j\in S}v_{j}=0
\end{equation}

We can note that if \textbf{v} is an eigenvector associated to 0 eigenvalue of $A+I$ matrix, \textbf{v} will also be an eigenvector associated to 0 eigenvalue of $(A+I)^{2}$, $(A+I)^{3}$, etc. So:

\begin{equation}
(A+I)^{2}\textbf{v}=0
\end{equation}

\begin{equation}
  \begin{pmatrix}
     2+\sum_{i=3}^{N}a_{1i}^{2} & 2+\sum_{i=3}^{N}a_{1i}^{2} & \cdots & 3a_{1N}+\sum_{i=3}^{N-1} a_{1i}a_{iN} \\
     2+\sum_{i=3}^{N}a_{1i}^{2} & 2+\sum_{i=3}^{N}a_{1i}^{2} & \cdots & 3a_{1N}+\sum_{i=3}^{N-1} a_{1i}a_{iN} \\
     \vdots  & \vdots  & \ddots & \vdots  \\
     3a_{1N}+\sum_{i=3}^{N-1} a_{3i}a_{iN} & 3a_{1N}+\sum_{i=3}^{N-1} a_{1i}a_{iN} & \cdots & 1+a_{1N}^2+\sum_{i=3}^{N-1}a_{iN}^{2}  
  \end{pmatrix}
   \begin{pmatrix}
  v_{1} \\
  v_{2} \\
  \vdots \\
  v_{N}
 \end{pmatrix}
 =0
\label{EQ2}
\end{equation}

The two first rows enable to write:

\begin{equation}
(2+\sum_{i=3}^{N}a_{1i}^{2})(v_{1}+v_{2})+\sum_{i=3}^{N}((3a_{1i}+\sum_{j=3,j\neq i}^{N} a_{1j}a_{ji})v_{i})=0
\label{A2}
\end{equation}

It is obvious that the trivial case $a_{13}=a_{14}=...=a_{1N}=0$ lead to the equation (\ref{Relation}).
The trivial case corresponds to a isolated complete subgraph.
By using the equations (\ref{A}) and (\ref{A2}), we obtain the following relations:

\begin{equation}
\left\{
    \begin{array}{ll}
	v_{1}+v_{2}=(2+\sum_{i=3}^{N}a_{1i}^{2})(v_{1}+v_{2}) \\
	\sum_{i=3}^{N}a_{1i}v_{i}=\sum_{i=3}^{N}((3a_{1i}+\sum_{j=3,j\neq i}^{N} a_{1j}a_{ji})v_{i}) \\
    \end{array}
\right. 
\label{SYST1}
\end{equation}

\begin{equation}
\left\{
    \begin{array}{ll}
	1+\sum_{i=3}^{N}a_{1i}^{2}=0 \\
	2a_{1i}+\sum_{j=3,j\neq i}^{N} a_{1j}a_{ji}=0 \\
    \end{array}
\right. 
\label{SYST2}
\end{equation}

However, in the non-trivial case, $1+\sum_{i=3}^{N}a_{1i}^{2}\neq 0$ and $2a_{1i}+\sum_{j=3,j\neq i}^{N} a_{1j}a_{ji}\neq 0$. So, the system (\ref{SYST2}) is absurd.
That leads to:

\begin{equation}
\left\{
    \begin{array}{ll}
	v_{1}+v_{2}=0 \\
	v_{3}=v_{4}=...=v_{N}=0 \\
    \end{array}
\right. 
\label{SYST3}
\end{equation}

As a result, the equation (\ref{Relation}) is verified.

\section{Eq (\ref{Relation}) and (\ref{Relation2}) through row equivalence and Gaussian elimination}

As we have written in the manuscript, the conditions (ii) and (iii) make the rank of $A-\lambda I$ decrease. More particularly, if $r$ is the number of linear combinations of rows and $N$ the size of the matrix, one obtains rank$(A-\lambda I)$=$N-r$. Mathematically, it results from the rank-nullity theorem according to which:

\begin{equation}
    \mbox{rank}(A-\lambda I)+\mbox{null}(A-\lambda I)=N \\
\end{equation}

with null the dimension of null space of $A-\lambda I$. Knowing that, it seems likely that we can take advantage of the row echelon form using Gaussian elimination. Indeed, we will see that it sheds light on Eq. \ref{Relation} and \ref{Relation2}. Before starting, we should recall that the three elementary row operations (swap the positions of two rows, multiply a row by a nonzero scalar and add to one row a scalar multiple of another) do not affect the solution set if the matrix is associated to a system of linear equations. 

\vspace{0.5cm}

First, we consider the condition (ii) through a simple but typical example. We take a duplicate node structure such as two nodes have the same neighbor. This kind of graph makes appear one 0 eigenvalue. In this case, we work on $A-\lambda I=A-0.I=A$ where $R_{1}=R_{2}$.

\begin{equation}
A=
 \begin{pmatrix}
  0 & 0 & 1 \\
  0 & 0 & 1 \\
  1 & 1 & 0  
 \end{pmatrix}
 \end{equation}
 
 Let us apply $L_{1} \leftrightarrow L_{3}$ and $L'_{3} \leftarrow L'_{3}-L_{2}$:
 
 \begin{equation}
A  \sim 
 \begin{pmatrix}
  1 & 1 & 0 \\
  0 & 0 & 0 \\
  0  & 0  & 1  
 \end{pmatrix}
\label{A+I}
 \end{equation}
 
 The previous matrix is a row echelon form. Since one row has all its entries equal to $0$, it clearly shows that rank$(A)=2$ whereas $N=3$. Let us come back to the eigen-equation with this equivalent form:
 
  \begin{equation} 
 \begin{pmatrix}
  1 & 1 & 0 \\
  0 & 0 & 0 \\
  0  & 0  & 1  
 \end{pmatrix}
  \begin{pmatrix}
  v_{1}  \\
  v_{2}  \\
  v_{3}   
 \end{pmatrix}
 =
   \begin{pmatrix}
  0  \\
  0  \\
  0   
 \end{pmatrix}
 \end{equation}
 
 It follows that: $v_{1}+v_{2}=0$ and $v_{3}=0$. So Eq. \ref{Relation} is verified.
 
\vspace{0.5cm}
 
 Now, we focus on the condition (iii) using the example reported by Figure 1 (b) of the manuscript for which $R_{1}+R_{4}=R_{2}+R_{5}$. It results one 1 eigenvalue and that is why we are going to work on $A+I$.
 
 \begin{equation}
A+I=
 \begin{pmatrix}
  1 & 1 & 0 & 0 & 0 \\
  1 & 1 & 1 & 0 & 0 \\ 
  0 & 1 & 1 & 1 & 0\\
  0 & 0 & 1 & 1 & 1 \\
  0 & 0 & 0 & 1 & 1 \\ 
 \end{pmatrix}
 \end{equation}
 
 Let us apply once more the Gaussian elimination. One obtains:
 
 \begin{equation}
A+I \sim
 \begin{pmatrix}
  1 & 0 & 0 & 0 & 1 \\
  0 & 1 & 0 & 0 & -1 \\ 
  0 & 0 & 1 & 0 & 0\\
  0 & 0 & 0 & 1 & 1 \\
  0 & 0 & 0 & 0 & 0 \\ 
 \end{pmatrix}
 \end{equation}
 
 As it was the case in the previous section, this row equivalent matrix enables to find that rank$(A+I)=4$ whereas $N=5$. Let us consider the eigen-equation with this equivalent form:
 
  \begin{equation}
 \begin{pmatrix}
  1 & 0 & 0 & 0 & 1 \\
  0 & 1 & 0 & 0 & -1 \\ 
  0 & 0 & 1 & 0 & 0\\
  0 & 0 & 0 & 1 & 1 \\
  0 & 0 & 0 & 0 & 0 \\ 
 \end{pmatrix}
  \begin{pmatrix}
  v_{1} \\
  v_{2} \\ 
  v_{3} \\
  v_{4} \\
  v_{5} \\ 
 \end{pmatrix}
 =
   \begin{pmatrix}
  0 \\
  0 \\ 
  0 \\
  0 \\
  0 \\ 
 \end{pmatrix}
 \end{equation}

It follows that $v_{1}+v_{2}+v_{4}+v_{5}=0$ and $v_{3}=0$. We recover Eq. \ref{Relation2}.

\section{Application examples of Eqs. (\ref{Relation}) and (\ref{Relation2})}

Since we have understood why Eq. (\ref{Relation}) and (\ref{Relation2}) occur, let us apply them to other examples.\\

First, let us take an example to illustrate the 
relation (\ref{Relation}): only two rows of an adjacency matrix verify the condition (ii) such as $R_{1}=R_{2}$ and 
all the other ones are linearly independent. Then, Eq. \ref{Relation} tells us that $v_{1,2} \neq 0$, 
$v_{1}+v_{2}=0$ and $v_{i}=0$ for $i \neq 1,2$ in every eigenvector associated to the 
studied degenerate eigenvalue. \\

Then, let us take an example for the relation (\ref{Relation2}): let us assume that an adjacency matrix satisfies the 
relation $R_{1}+R_{2}=R_{3}+R_{4}$ and all the other rows are linearly independent. Then, Eq. \ref{Relation2} shows that $v_{1,2,3,4} \neq 0$, $v_{1}+v_{2}+v_{3}+v_{4}=0$ and $v_{i}=0$ for $i \neq 1,2,3,4$ in every eigenvector associated to the studied degenerate eigenvalue. In other words, this means that the rows $i$ with $v_i \neq 0$ are linearly
dependent if and only if they belong to a same structure, namely a same linear combination of rows. \\

In the case where there are two different linear combinations of rows as follows $R_{1}+R_{2}=R_{3}$ and $R_{4}+R_{5}=R_{6}$, so $v_{1,2,3,4,5,6} \neq 0$, $v_{1}+v_{2}+v_{3}=0$, $v_{4}+v_{5}+v_{6}=0$ and $v_{i}=0$ for $i \neq 1,2,...,6$.